\documentclass[twocolumn,aps]{revtex4}
\usepackage{epsfig}
\begin{document}
\title{
Resolving singular forces in cavity flow: Multiscale
modeling from atoms to millimeters
}
\author{Xiaobo Nie$^{1}$, Mark. O. Robbins$^{1,2}$,
and Shiyi Chen$^{2,3}$}

\affiliation{
${}^{1}$Department of Physics and Astronomy, The Johns Hopkins
University, Baltimore, MD 21218 \\
${}^{2}$Department of Mechanical Engineering, The Johns Hopkins
University, Baltimore, MD 21218\\
${}^{3}$CCSE and SKLTCS, Peking University, Beijing, China }

\begin{abstract}
A multiscale approach for fluid flow is developed that retains an atomistic
description in key regions.  The method is applied to a classic
problem where all scales contribute: The force on a moving wall bounding a
fluid-filled cavity.
Continuum equations predict an infinite force due to stress singularities.
Following the stress over more than six decades in length in systems with
characteristic scales of millimeters and milliseconds allows us to resolve
the singularities and determine the force for the first time.
The speedup over pure atomistic calculations is more than fourteen
orders of magnitude. We find a universal dependence on the macroscopic
Reynolds number, and large atomistic effects that
depend on wall velocity and interactions.
\vspace{2pt}
\end{abstract}
 \maketitle

Processes that span a wide range of length scales pose profound
theoretical challenges \cite{glimm97,hou05}.
The different length scales must be followed with
different time resolutions and may require qualitatively different
descriptions of matter.
For example, discrete atomistic effects may be important in regions of
high stress or rapid spatial variation,
while other regions are most naturally
and efficiently modeled as a continuous medium.
Important examples of such problems include adhesion and friction
\cite{urbakh04,luan05},
deformation of crystalline solids \cite{uchic04,weiss02},
distribution and flow of charges at biological interfaces,
flow near solid surfaces \cite{thompson97}, and the many
cases where continuum equations lead to singularities.

Several innovative paradigms for bridging between atomistic and
macroscopic scales have been proposed in recent years,
and tested against purely atomistic simulations in small
idealized systems
\cite{li98,nie04,connell95,hadjiconstantinou97,flekkoy00,shenoy98,ren05,kevrekidis04}.
A few have been applied to specific problems with a large range of scales.
Notable examples include calculations of crack propagation
in silicon that include electronic structure near the crack
tip \cite{broughton99},
and calculations of indentation with the quasicontinuum
method \cite{knap03}.
However, these applications have only reached micrometer
length and nanosecond time scales, and the main effect of large
scales is to provide appropriate boundary conditions for the
atomistic region.

In this paper we consider a classic problem in fluid mechanics
where all length scales contribute equally: the force on a
moving boundary of a fluid filled cavity.
Using different spatial and temporal resolutions in different
regions allows us to treat cavities with dimensions of
order millimeters
and characteristic times of tens of milliseconds.
Our multiscale approach accelerates the calculation by more
than fourteen orders of magnitude compared to brute force
atomistic simulations.
Spanning many decades in length scale allows us to build a simple
scaling relation for the total force $F$ that captures both
atomistic effects and the influence of the only parameter in continuum
theory, Reynolds number.

\begin{figure}
 \bigskip
  \centerline{ {\epsfig{file=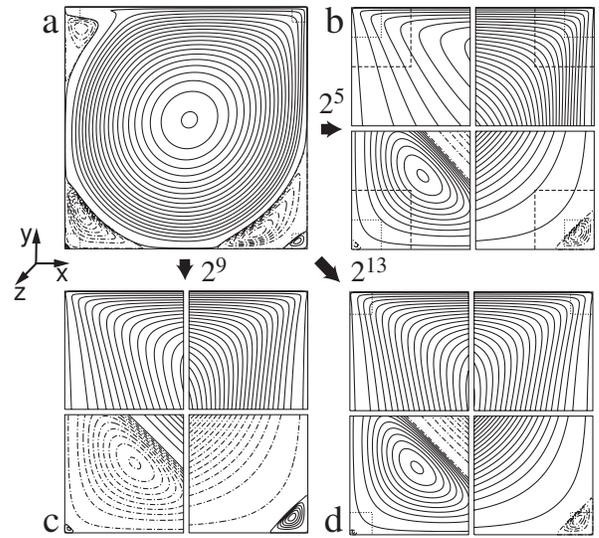,width=220pt}} }
  \caption{Geometry and streamlines for cavity flow at
  $Re \equiv \rho U L/\mu=6400$.
  a) The square cavity has edge $L=10^6\sigma \sim 0.3$mm
  in the $x-y$ plane,
  %=2.49 \times 10^5 \sigma$
  the top wall moves to the right at speed $U$,
  the flow is independent of $z$, and the natural time scale
  to reach steady state flow is $\sim 40$ms.
  Solid (dash-dotted) streamlines indicate clockwise (counter-clockwise) flow.
  In (b-d), flow near each corner is magnified by the indicated factor.
  The coupling of different resolutions is illustrated in b.
  The coarser solution provides boundary conditions along the
  outer boundary of the finer grid (dashed lines) and
  obtains them along the dotted lines.
  In (d), dotted lines indicate the boundaries of the $\sim 12 \sigma$ wide
  regions that are treated atomistically.
  }\label{fig:flow}
\end{figure}

Cavity flow has intrigued scientists because of singularities
that can not be resolved by purely continuum methods.
Figure \ref{fig:flow}a illustrates the cavity geometry.
The top wall is displaced to the right at fixed velocity ${\bf U}$
and the other walls are stationary.
The traditional continuum approach to this problem
uses the Navier-Stokes (NS) equations and no-slip boundary
conditions at the walls \cite{batchelor67,pan67,ghia82}.
The no-slip condition requires that the fluid velocity ${\bf u}$
vanishes near static walls, and equals ${\bf U}$ at the moving wall.
The discontinuity in boundary condition at corners between
moving and fixed walls causes the stress to diverge
as the inverse of the distance from the corner, $r$.
The total force on the wall is the integral of the stress,
and diverges logarithmically in continuum theory.

Koplik and Banavar pioneered the use of molecular dynamics (MD)
to study the singularities in cavity flow \cite{koplik95b}.
They observed a breakdown of the no-slip boundary condition
within atomic distances from the corner, and a corresponding saturation
of the stress.
Similar effects cut off stress singularities in the
closely related problem of spreading fluids \cite{koplik88,thompson89}.
Koplik and Banavar's purely atomistic approach limited the cavity
length to $L=17 \sigma$, where $\sigma \sim 0.3$nm corresponds
to a molecular diameter.
We recently extended $L$ to $250\sigma \sim 75$nm using a hybrid method
that treated singular regions atomistically and the remainder
of the cavity as a continuum.
This approach allowed us to analyze the breakdown of the
continuum boundary conditions over a wider range of $U$,
and to determine its microscopic origins \cite{nie04b}.
The stress deviates from the singular continuum
solution for $r<S(U)$.
At low $U$, the intrinsic discreteness of atomic fluids leads
to $S \sim \sigma$.
At large $U$, the interfacial stress is high enough to produce
non-Newtonian effects \cite{thompson97}, and $S$ rises linearly with $U$.

To study cavities with millimeter scale dimensions, two major
improvements must be made on previous work.
The first is to vary the resolution in the continuum region
to efficiently describe the rapid increase in velocity gradients
near the singular corners.
We chose to do this using a local refinement approach
\cite{mccormick89}.
The second is to span the wide range of time scales associated
with different spatial resolutions.
While the motion of atoms near the corner must be followed with
time steps of order $10^{-14}$s, the time for flows to equilibrate
on millimeter scales, $\sim 100 L/U$, may be of order seconds.
To overcome this obstacle we use the optimum time step to
obtain the steady state flow at each spatial scale, and then
enforce self-consistency between scales.

The details of our approach and a sample flow field
are illustrated in Fig. \ref{fig:flow}.
At the continuum scale the flow is independent of $z$,
and satisfies the NS equations with vicosity $\mu$,
fixed density $\rho$, and no-slip boundary conditions.
At each scale the NS equations are discretized on a square grid
of cells with width $h$, and
the steady state flow is obtained
using an artificial compressibility method \cite{peyret83}
with time step of 1/4 -- 1/2 $h/U$.
On the coarsest scale $h=L/256$.
This resolution is inadequate near the corners,
where $h$ is decreased by successive factors of two using
an iterative refinement scheme.

At each stage of the iteration, solutions at two
resolutions provide boundary conditions for each other.
The geometry is illustrated in Fig. \ref{fig:flow}b.
Both resolutions use a 64 by 64 array of square cells.
The finer grid lies in the inner quarter of the coarse grid (dashed lines),
and receives boundary conditions on its outer edge.
It in turn provides boundary conditions for the coarser grid
along the dotted lines.
The overlap region between dashed and dotted lines
prevents discontinuities due to sudden changes in resolution
\cite{nie04,connell95,flekkoy00}.

This refinement scheme is iterated until the overlap region
reaches nanometer scales.
There the finest resolution results are obtained from
MD simulations that can be extended all the way into the corner
(Fig. \ref{fig:flow}d).
At the outer edge of the MD region the mean atomic velocity is constrained
to follow the finest continuum solution ($h=0.95\sigma$)
and particles are added or removed to match the continuum
flux \cite{nie04,nie04b}.
Average MD velocities provide boundary conditions for the
continuum solution along an inner square whose edge is 6 cells long.
A global steady state solution is obtained by iterating from
coarsest to finest scale and then back to the coarsest
scale until all boundary conditions are consistent.
This typically requires ten to twenty iterations, depending
on the desired accuracy.

To obtain a smooth solution, the continuum model must accurately
describe the atomistic behavior at the outer boundary of the MD region.
This requires consistent choices of $\mu$ and molecular interactions.
Following previous work \cite{koplik95b,nie04b}, we consider
fluid atoms of mass $m$ interacting with a Lennard-Jones (LJ)
potential of characteristic energy $\epsilon$ and diameter
$\sigma$.
The potential is truncated at $r_c=2.2 \sigma$,
and the mass density $\rho=0.81 m \sigma^{-3}$.
The geometry and interactions of the crystalline walls
are chosen to produce a no-slip boundary condition far from
the corner \cite{nie04b}.
Discrete wall atoms are on the sites of a (111)
surface of an fcc crystal of lattice constant 1.204 $\sigma$
and interact with the fluid with a LJ potential with
energy $\epsilon_{wf}=0.95 \epsilon$.

Within the MD region, the motion of particles is fully three-dimensional.
However the mean flow velocities are independent of $z$,
and periodic boundary conditions with period $L_z$ are applied
in this direction.
The equations of motion are integrated using the Verlet scheme with
time step $\Delta t_{MD} = 0.005 t_{LJ} $, where $t_{LJ} \equiv
(m{\sigma}^2/\epsilon)^{1/2}$ is the characteristic time of the
LJ potential. A constant temperature $k_B T = 1.1 \epsilon$
is maintained by applying a Langevin thermostat\cite{grest86}
with damping rate $\Gamma = 1 t_{LJ}^{-1}$
in the $z$ direction.
The dynamic viscosity\cite{connell95} of the LJ fluid $\mu = 2.14\epsilon
t_{LJ} {\sigma}^{-3}$ is used in the NS equations.

Two lengths characterize transitions in flow behavior near each corner.
Inertial effects are significant for $r > r_I \equiv \mu/\rho U$,
while deviations from continuum behavior occur for $r < S(U)$.
At intermediate scales our results follow the analytic solution
for Stokes (non-inertial) flow \cite{batchelor67}.
Near the top corners the streamlines are scale invariant
because ${\bf u}$ only depends on the angle relative to the
moving wall.
The stress diverges as $1/r$ since ${\bf u}$ changes by $U$
over a length of order $r$.
A series of counter-rotating vortices forms near the bottom corners.
The change in scale between panels in Fig. \ref{fig:flow} was
chosen to illustrate the predicted self-similarity under
$r \rightarrow r/16.4$ and a change in direction of rotation
\cite{moffatt64,pan67}.
The vortices are cut off at $S(0) \sim \sigma $ by atomistic effects,
and this is responsible for the small deviation between panels c and d.

\begin{figure}
\bigskip
  \centerline{ {\psfig{file=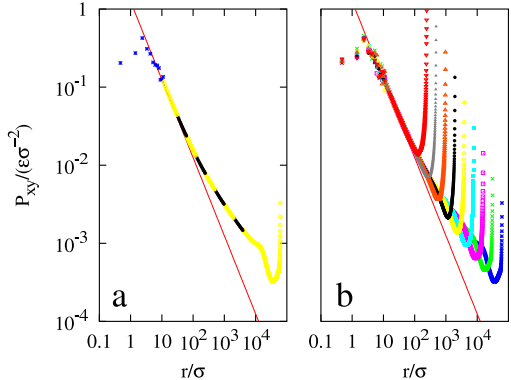,width=220pt}} }
  \caption{Shear stress as a function of distance $r$ from the upper
  left corner for $U=0.27 \sigma/t_{LJ}$.
  Qualitatively similar scaling occurs near the upper-right corner.
  In (a) $Re=6400$, $L=6.25 \times 10^4 \sigma$, $r_I=9.8\sigma$,
  $S\approx 2.2\sigma$, lines of
  alternating color indicate continuum results from successive resolutions,
  and asterisks show MD results.
  Results for $Re=25$ to $6400$ obtained by increasing $L$ by factors
  of two are compared in (b).
}\label{fig:stress}
\end{figure}

Fig.~\ref{fig:stress}(a) shows a plot of
the shear stress $\tau$ along the moving wall
at Reynolds number $Re \equiv \rho U L/\mu=6400$
as a function of distance $r$ from the upper-left corner.
Alternating colors are used for data from different resolutions to illustrate
the smooth matching.  The asterisks are from the force
per unit area on wall atoms in the atomistic region.
%MD simulations.
At intermediate scales, the stress follows the divergence
predicted by the Stokes solution
$\tau = (4\pi)/(\pi^2-4) (\mu U)/r$ (straight line).
At large scales the stress decays more slowly than $1/r$,
and at $r<S\approx 2.2\sigma$ the stress singularity is cutoff by atomistic
effects.
Fig.~\ref{fig:stress}(b) shows that changing $Re$
by increasing $L$ only changes the flow in the outer region.
At scales of order $L/2$ the other corners become important,
but for $r \ll L/2$ all results fall onto a common curve.
This shows that $S$ is independent of $L$
and only depends on $U$ and atomic properties.

\begin{figure}
\bigskip
  \centerline{ {\psfig{file=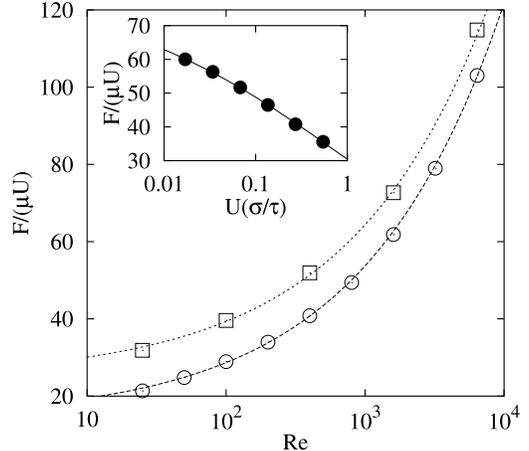,width=220pt}} }
  \caption{Dimensionless force per unit length on the sliding wall $F/\mu U$
  as a function of Reynolds number $Re=\mu U L/\rho$.
  Results for $U=0.27 \sigma/t_{LJ}$ (circles) and $U=0.068\sigma/t_{LJ}$
 (squares) follow Eq. \ref{eq:fit} (dashed lines) with $S=2.2$ and
 $0.55$ respectively.
 The inset compares the calculated force vs. velocity at $Re=400$ (circles)
 to Eq. \ref{eq:fit} (line).
  }\label{fig:force}
\end{figure}

The total shear force on the moving wall is an integral
of the shear stress along it.
Fig. \ref{fig:force} shows the force per unit
length along the $z-$direction, $F$, as a function of $Re$.
If the Stokes solution applied, each factor of two in length
scale would contribute the same force and the total would
diverge.
This divergence is cutoff by atomistic effects at $r<S$
but inertial effects enhance the force from $r>r_I$.
To determine $F$, one must resolve the inertial
effects between $r_I$ and $L=Re\  r_I$ as well as the atomistic
behavior at the corner.
Our approach enables us to span
this wide range of scales ($> 10^5$) for the first time.

If atomistic effects were not important, the dimensionless
force on the wall $f \equiv F/\mu U$ would only depend on $Re$.
However, Fig. \ref{fig:force} shows that changing $U$ at fixed
$Re$ produces large shifts in $f$.
A complete description of $f$ can be obtained by considering the
separate contributions, $f_i$, of the three
scaling regimes in the stress shown in Fig. \ref{fig:stress}.
They are well-separated as long as $Re=L/r_I \gg 1$ and the
Reynolds number at molecular scales $Re_m \equiv S/r_I \ll 1$.

The dependence on $Re$ comes entirely from the outer region $r>r_I$,
which is why curves for different $U$ in Fig. \ref{fig:force}
are related by a constant shift.
As shown in Fig. \ref{fig:stress}(b), increasing $Re$
extends the range of the inertial tail that contributes
to the dimensionless force, causing
$f_3(Re)$ to rise with $Re$.
For $Re > 10$ our numerical results are well described by
$f_3 = a+ b {Re}^\alpha$
with $a=3.85$, $b=1.98$ and $\alpha=0.434$.

In the inner region near each corner, $r<S$, the
stress is determined entirely by $U$ and atomic properties,
such as $t_{LJ}$, $\sigma$, $\rho$, and wall geometry
(Fig. \ref{fig:stress}(b)).
For fixed interactions, we can write the contribution from
this regime as $f_1(Ut_{LJ}/\sigma)$.
In fact we find $f_1$ has a constant value of about 4.3.
The reason is that the range of integration grows linearly
with $S$ while the stress scales inversely with $S$.
Indeed assuming the stress for $r<S$ is equal
to the Stokes stress at $S$ yields $f_1=8\pi/(\pi^2-4) \approx 4.28$,
which is consistent with the numerical results.
The value of $f_1$ is as much as 20\% of the total wall force
in Fig. \ref{fig:force}, yet it comes from an inner region
that is never larger than a few molecular diameters.
This is clear evidence of the direct effect of atomic scale
behavior on the macroscopic force.

In the intermediate region, $S<r<r_I$, the Stokes solution applies,
and can be integrated to give $f_2(r_I/S)=8\pi/(\pi^2-4) \ln(r_I/S)$.
This term is responsible for the velocity dependence in the
total dimensionless force:
\begin{equation}
    \frac{F}{\mu U}=f_1+\frac{8\pi}{\pi^2-4} \ln(r_I/S)
    + a+b{Re}^{\alpha} \ \ .
    \label{eq:fit}
\end{equation}
The ratio $S/r_I$ reflects the range at which atomistic
effects cut off the Stokes region, and decreases as $U$ decreases.
The inset to Fig. \ref{fig:force} shows that numerical results
for the velocity dependence of $F$ are consistent with
$S=S_0 + k Ut_{LJ}$ with $S_0=0.3\sigma$ and $k=7$.
This form for $S$ is consistent with our previous studies \cite{nie04b},
which showed $S$ approached a constant of atomic scale
at low velocities and rose linearly with $U$ at
high velocities due to non-Newtonian effects.

Our multiscale method has allowed us to span the wide range
of length scales ($> 10^5$) that contribute to the drag force on
the moving wall of an ideal cavity, and to extract a simple and accurate
physical description (Eq. \ref{eq:fit}) of the important
contributions from each scale.
We have used it to treat cavities with dimensions of
order millimeters and natural time scales approaching
seconds, and the approach is readily extended to still
larger scales.
Its major limitation is that it assumes a steady state solution,
while cavity
flow becomes turbulent at $Re > 8000$ \cite{cazemier98}.
It may be possible to extend simulations into the
turbulent regime using ideas like the ``equation-free''
approach of Kevrekidis et al.\cite{kevrekidis04}.
Turbulent fluctuations occur on the longer time scales associated
with coarser resolutions, and the finer scales can be iterated to
a quasistatic state that follows the coarse solution.
This would allow calculations with the same coarse time step that
is used in a completely continuum description, while retaining the
crucial atomic detail.
We hope that our work encourages such efforts, as well
as applications to other important multiscale problems
such as contact-line motion and contact mechanics.

{\bf Acknowledgments.} This material is based upon work
supported by the U.S. National Science Foundation under Grant No.
CMS-0103408 and the Digital Laboratory for Multiscale Science.

\end{document}